\documentclass[letterpaper]{article} 
\usepackage{aaai2027}  
\usepackage[hyphens]{url}  
\usepackage{graphicx} 
\usepackage{natbib}  
\usepackage{caption} 
\usepackage{amsmath}
\usepackage{amssymb}
\usepackage{algorithm}
\usepackage{algorithmic}

\usepackage{newfloat}
\usepackage{listings}
\DeclareCaptionStyle{ruled}{labelfont=normalfont,labelsep=colon,strut=off} 
\floatstyle{ruled}
\newfloat{listing}{tb}{lst}{}
\floatname{listing}{Listing}

\usepackage{booktabs}
\usepackage{multirow}
\usepackage{array}

\title{In-Context Adapter Generation for Compositional Generalization in Large Audio Language Models}
\author{Yiwei Guo$^{1,2}$, Siqi Huang$^{3}$, Bohan Li$^{1,2}$, Jing Peng$^{1,2}$, Hankun Wang$^{1,2}$, Zhihan Li$^{1,2}$, Kai Yu$^{1,2}$}
\affiliations{$^1$X-LANCE Lab, MoE Key Lab of Artificial Intelligence, Shanghai Jiao Tong University\\$^2$Jiangsu Key Lab of Language Computing\\$^3$Xi'an Jiaotong University}

\usepackage{colortbl}
\usepackage{tcolorbox}
\usepackage{tabularx}
\usepackage{ragged2e}
\tcbuselibrary{skins,breakable,listings}
\definecolor{promptbg}{gray}{0.96}
\definecolor{promptframe}{gray}{0.55}
\newtcblisting{llmprompt}{%
	colback=promptbg,
	colframe=promptframe,
	boxrule=0.6pt,
	arc=5pt,
	left=8pt,
	right=8pt,
	top=6pt,
	bottom=6pt,
	boxsep=2pt,
	listing only,
	listing options={%
		basicstyle=\footnotesize\ttfamily,
		breaklines=true,
		columns=fullflexible,
		keepspaces=true,
		showstringspaces=false,
	},
}
\definecolor{tablegray}{gray}{0.94}
\newcommand{\graycell}[1]{\cellcolor{tablegray}#1}
\newcommand{\stackcell}[1]{\parbox[c]{\linewidth}{\raggedright #1}}
\newcommand{\taskclass}[2]{\raisebox{-0.7ex}{\parbox{0.12\textwidth}{\raggedright #1\\{\footnotesize #2}}}}
\newcolumntype{L}[1]{>{\raggedright\arraybackslash}m{#1}}
\newcolumntype{C}[1]{>{\Centering\arraybackslash}p{#1}}

\usepackage[hidelinks,breaklinks]{hyperref}

\begin{document}

\maketitle

\begin{abstract}
Large audio language models (LALMs) have achieved strong performance on standard speech and audio understanding tasks, yet their ability to systematically generalize to novel task compositions remains underexplored.
Taking a meta-learning view, we extend Text-to-LoRA (T2L) to LALMs, generating instruction-conditioned LoRA adapters with a Transformer hypernetwork rather than learning a single fixed adapter shared across instructions.
To exploit fine-grained task information, we introduce in-context adapter generation (ICAG), a token-level hypernetwork architecture; we call the resulting method T2L-ICAG.
We evaluate this approach on two representative LALMs using SIFT-50M for supervised fine-tuning and construct a controlled compositional evaluation suite covering five task classes.
Across two LALM backbones, T2L-ICAG consistently improves over static LoRA and also the global embedding-based T2L baseline. 
Coverage and capacity analyses further show that these gains are larger on structurally out-of-domain tasks and cannot be explained by parameter count alone.
These results suggest that generating task-specific adapters from instructions is a promising direction for improving compositional generalization of LALMs.
\end{abstract}


\section{Introduction}
Large audio language models (LALMs) have achieved strong performance in general audio and speech understanding tasks, showing impressive perception and reasoning abilities~\cite{peng2026survey,luo2026surveylargeaudiolanguage}.
However, it remains unclear whether LALMs can systematically recombine familiar capabilities and task schemas into novel task compositions without direct training examples of the target compositions.
This raises the question of \textbf{compositional generalization}, an important property of true intelligence.

Compositional generalization requires a model to transfer knowledge from familiar task components to novel task compositions~\cite{Fodor1988ConnectionismAC,lake2018generalization}.
This capability is especially important for audio and speech understanding, where acoustic functionalities, such as gender perception, language identification and emotion recognition, are separable from task schemas.
Here, a task schema refers to the structural form in which one or more acoustic functionalities are invoked over one or more audio inputs, such as single-input recognition, two-input comparison, selective execution, sequential execution, and so on.
Together, these functionalities and schemas span a large compositional task space.
LALMs with compositional generalization ability should generalize to unseen task compositions built from familiar components, even when those compositions are not explicitly observed during training.
For example, a model that has learned \textit{gender recognition}, \textit{duration comparison} and \textit{pitch comparison} should be able to understand comparison as a reusable task schema and naturally perform \textit{gender comparison}.
However, we find that existing open-source LALMs struggle with such compositional transfer, and the literature lacks systematic studies of this problem in the audio domain.

The problem of compositional generalization is not new and has long been studied in natural language processing~\cite{Fodor1988ConnectionismAC}.
Although scaling LLMs has partially alleviated it~\cite{yang-etal-2024-exploring,sakai-etal-2025-revisiting}, audio introduces an additional cross-modal grounding challenge.
Successful composition requires a LALM not only to interpret a novel instruction structure, but also to bind each requested functionality to the appropriate acoustic evidence and audio input.
A model may therefore recognize gender or emotion in isolation yet fail when the same capability must be selected, ordered or applied across multiple inputs.
Exhaustively enumerating task compositions during training may alleviate this problem, but is expensive and does not reflect human-like generalization.
This motivates us to explore adaptation methods that generalize beyond explicit task coverage.

In modern LLMs, teaching a pretrained model to perform new tasks is usually done through supervised fine-tuning (SFT), often with low-rank adaptation (LoRA)~\cite{hu2022lora}.
By decomposing parameter updates into low-rank matrices, LoRA provides efficient task adaptation with a small number of trainable parameters.
However, a static LoRA adapter still stores adaptation knowledge in a fixed set of parameters optimized over the observed training task space.
This perspective motivates a meta-learning view: instead of only learning task-specific parameters, the model can learn how adaptation should be performed for a new task.
Hypernetworks provide a natural realization of this idea by generating model parameters conditioned on task information.

In this study, we extend Text-to-LoRA (T2L)~\cite{text2lora} to LALMs to improve compositional generalization.
Instead of training a static LoRA adapter, T2L generates instruction-conditioned LoRA weights with a hypernetwork.
We propose in-context adapter generation (ICAG), a bidirectional Transformer hypernetwork architecture, and call the resulting method T2L-ICAG.
In this way, adapters are generated with fine-grained task-specific control from instruction tokens rather than from a single global task semantic.
The LoRA parameters no longer need to be directly distributed over the observed training task space; instead, the hypernetwork can condition the generated adapters on each task description, providing a flexible parameterization for different combinations of acoustic skills and task schemas.
Through ICAG, the hypernetwork attends to each instruction token when generating adapter weights, which preserves fine-grained instruction information when generating adapters for unseen task compositions.

We experiment with Kimi-Audio~\cite{ding2025kimi} and Qwen2.5-Omni~\cite{xu2025qwen25omnitechnicalreport} on the diverse instruction tuning dataset SIFT-50M~\cite{pandey2025sift}.
The results show that generative LoRA adaptation significantly improves performance on unseen task compositions compared to static LoRA at the same rank.
T2L-ICAG also outperforms T2L-emb, the global embedding-based architecture in \citet{text2lora}.
In summary, our contributions are:
\begin{itemize}
	\item To the best of our knowledge, we present the first empirical study of compositional generalization in LALMs.
	\item We extend T2L-based adapter generation to LALMs and propose ICAG, a token-level Transformer hypernetwork architecture that generates instruction-conditioned LoRA adapters.
	\item Across two LALM backbones, T2L-ICAG outperforms static LoRA and T2L-emb, with larger gains on structurally novel tasks that cannot be explained by training-data overlap or parameter count alone.
\end{itemize}

\section{Background}

\subsection{Large Audio Language Models}
In this work, we consider LALMs as multimodal language models that can understand both audio and text inputs, i.e. speech and audio understanding models.
Prevalent LALMs usually consist of an audio encoder and a decoder-only LLM.
Through scaling in representation alignment, supervised fine-tuning and post-training, researchers have established LALMs as a unified multi-task audio and speech understanding model capable of recognition, translation, audio captioning, reasoning and more~\cite{peng2026survey,luo2026surveylargeaudiolanguage}.
Well-known LALMs include Qwen2-Audio~\cite{chu2024qwen2audiotechnicalreport}, Kimi-Audio~\cite{ding2025kimi}, DeSTA2.5-Audio~\cite{DESTA}, etc.
Current omnimodal models also possess generic audio understanding abilities, e.g. Qwen Omni series~\cite{xu2025qwen25omnitechnicalreport,qwenteam2026qwen35omnitechnicalreport}.

\subsection{Compositional Generalization}
Compositional generalization, or {systematic generalization}, is the ability to compositionally process novel structures built from familiar components~\cite{Fodor1988ConnectionismAC}. 
While trivial for humans, compositional generalization remains a longstanding challenge for modern connectionist models.
\citet{lake2018generalization} formalized this challenge through the SCAN benchmark.
Subsequent works have extended this evaluation to broader settings \cite{COGS, keysers2020measuring,gSCAN}.
Despite remarkable scaling, recent evaluations suggest that compositional generalization remains imperfect~\cite{sakai-etal-2025-revisiting} in text LLMs.
However, compositional generalization remains largely underexplored for multimodal LALMs operating on audio and speech.
Note that complex audio processing abilities like captioning~\cite{ma2026omnicaptioner} and reasoning~\cite{xie2025audioreasonerimprovingreasoningcapability} do not necessarily demonstrate compositional generalization, since generalization concerns transfer from seen to unseen tasks, rather than task complexity.

\subsection{Meta-Learning and LoRA Hypernetworks}
Meta-learning~\cite{MAML} aims to extract inductive biases across tasks to enable rapid adaptation to unseen novel tasks with minimal data, and is believed to benefit compositional generalization for neural networks~\cite{lake_human-like_2023}.
Hypernetworks~\cite{ha2017hypernetworks} have emerged as a powerful architecture for dynamic parameter generation in model-based meta-learning.
In the LLM era, efforts have been made to design hypernetworks targeted at generating LoRA weights for modularized adaptation and adapter reuse, including HyperDecoders~\cite{ivison2022hyperdecoders}, Arrow routing~\cite{ostapenko2024towards}, Text-to-LoRA~\cite{text2lora}, ICM-Fusion~\cite{ICMfusion}, etc.
Similar LoRA hypernetworks have also been studied for context compression, like Doc-to-LoRA~\cite{charakorn2026doc2lora} and SHINE~\cite{liu2026shine}.
Despite these advances in text LLMs, the meta-learning view and hypernetwork-based adaptation remain largely underexplored for LALMs.

\section{Method}
\subsection{Formulation of LoRA Hypernetworks}
\begin{figure}[t]
    \centering
    \includegraphics[width=0.9\linewidth]{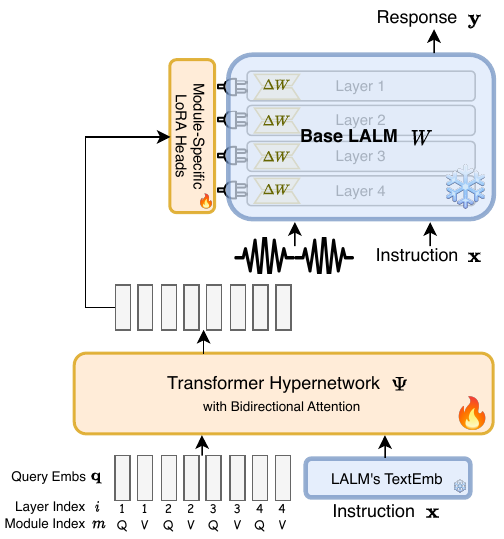}
    \caption{Overview of the proposed ICAG hypernetwork. An LALM with four layers and two LoRA modules (query and value projections) is used as an example.}
    \label{fig:overview}
\end{figure}

Given a base LALM with weights $\mathcal W$, after finetuning on a set of tasks, the model weights are updated by an adapter collection $\Delta\mathcal W$.
In the standard LoRA setting, each matrix update $\Delta W$ is decomposed into low-rank matrices $A$ and $B$.
Denote the number of layers as $L$ and the number of module types as $M$ (e.g. query projection, value projection, etc.).
For a linear weight matrix $W_{i,m}\in \mathbb R^{d_\text{out}\times d_\text{in}}$ at layer index $i \in [L]$ and module index $m \in [M]$, LoRA adapter is defined as 
\begin{equation}
	\Delta W_{i,m} = \frac{\alpha}{r} B^T_{i,m}A_{i,m}
	\label{eq:lora-update}
\end{equation}
where $B_{i,m} \in \mathbb R^{r\times d_\text{out}}$ and $A_{i,m} \in \mathbb R^{r\times d_\text{in}}$ are the low-rank matrices, $r$ is the rank of the LoRA adapter, and $\alpha$ is the LoRA scaling coefficient.
Suppose the training dataset is $\mathcal D=\{(\mathbf x_j, \mathbf a_j, \mathbf y_j)\}_{j=1}^{|\mathcal D|}$, where $\mathbf x_j$ is the text instruction sequence, $\mathbf a_j$ is the audio input sequence, and $\mathbf y_j$ is the answer sequence.
$\Delta \mathcal W$ is usually optimized by the SFT loss:
\begin{equation}
	\Delta\mathcal W^\star
	= \arg\min_{\Delta\mathcal W}
	\mathbb E_{(\mathbf x, \mathbf a, \mathbf y)\in \mathcal D}
	\mathcal L_{\text{SFT}}(\mathcal W+\Delta\mathcal W; \mathbf x, \mathbf a, \mathbf y).
\end{equation}

In the hypernetwork formulation, $\Delta W$ is generated by a hypernetwork $\Psi$ rather than directly optimized.
This hypernetwork takes as input the instruction sequence $\mathbf x$, layer index $i$ and module index $m$, and outputs the LoRA adapter $\Delta W_{i,m} = \Psi_{\boldsymbol\phi}(\mathbf x, i, m)$.
We use $\Psi_{\boldsymbol\phi}(\mathbf x)=\{\Delta W_{i,m}\mid i\in[L], m\in[M]\}$ to denote the full set of generated LoRA adapters for all target layers and modules, where $\boldsymbol\phi$ denotes the hypernetwork parameters.
Then, the hypernetwork is optimized with the same SFT criterion:
\begin{equation}
	\boldsymbol\phi^\star
	= \arg\min_{\boldsymbol\phi}
	\mathbb E_{(\mathbf x, \mathbf a, \mathbf y)\in \mathcal D}
	\mathcal L_{\text{SFT}}(\mathcal W+\Psi_{\boldsymbol\phi}(\mathbf x); \mathbf x, \mathbf a, \mathbf y).
\end{equation}
During training, gradients from the cross-entropy loss are backpropagated from the LoRA adapter weights to the hypernetwork parameters.

\subsection{Hypernetwork Architecture}

With hypernetwork $\Psi$, we can generate instance-specific LoRA adapters from the input instruction.
However, the architecture of the hypernetwork is not trivial.
Previous work, T2L~\cite{text2lora}, uses a global embedding-based approach: 
Let $\text{SentEmb}(\mathbf x)$ be a global sentence embedding for the instruction sequence $\mathbf x$ from a pretrained embedding model.
Then, $\Psi$ is formulated as a multi-layer perceptron $\Psi(\mathbf x,i,m) = \text{MLP}(\text{SentEmb}(\mathbf x), i, m)$.
While this exhibits generalization in the text domain with hundreds of training tasks, we find it suboptimal for LALMs.
Hence, we propose a Transformer-based hypernetwork architecture for in-context adapter generation (ICAG).

In our formulation, the instruction is tokenized by the LALM's tokenizer, embedded with the frozen token embedding matrix of the backbone model, and projected to the hidden dimension of the hypernetwork.
We denote the projected instruction token embeddings as $\text{TextEmb}(\mathbf x)\in\mathbb R^{|\mathbf x|\times d_h}$, where $|\mathbf x|$ is the number of instruction tokens and $d_h$ is the hypernetwork hidden size.
For $L$ decoder layers and $M$ target module types, we create one learnable query token for each layer-module pair, resulting in $LM$ query tokens in total.
For example, with two layers and two target modules (query and value projections), the prefix contains four query tokens whose layer-index sequence is $[1,1,2,2]$ and module-index sequence is $[\texttt{Q},\texttt{V},\texttt{Q},\texttt{V}]$.
Let $\mathbf q_{i,m}\in \mathbb R^{d_h}$ be the query token for the pair $(i,m)$, and $E_{\text{layer}}$ and $E_{\text{module}}$ be learnable index embedding tables.
The prefix token $\mathbf z_{i,m}$ for pair $(i,m)$ is
\begin{equation}
	\mathbf z_{i,m}=\mathbf q_{i,m}+E_{\text{layer}}(i)+E_{\text{module}}(m).
\end{equation}
We concatenate all prefix tokens with the instruction token embeddings and feed the sequence into a bidirectional Transformer encoder:
\begin{equation}
	[\mathbf z_{1,1};\ldots;\mathbf z_{L,M}; \text{TextEmb}(\mathbf x)] \in \mathbb R^{(LM + |\mathbf x|)\times d_h}.
\end{equation}
Let $H\in\mathbb R^{(LM+|\mathbf x|)\times d_h}$ be the output hidden states of the Transformer encoder.
We take the first $LM$ hidden states corresponding to the prefix tokens.
For each layer-module pair $(i,m)$, let $\mathbf h_{i,m}$ denote the corresponding prefix hidden state.
We then map $\mathbf h_{i,m}$ through a module-specific linear head to generate the LoRA matrices:
\begin{equation}
	[A_{i,m}, B_{i,m}] = \mathrm{Head}_m(\mathbf h_{i,m}).
\end{equation}
In practice, these heads output vectorized LoRA weights, which are then reshaped before being merged into the backbone.
During training, the LALM backbone and its input embedding matrix are frozen; only the hypernetwork parameters are updated through the standard SFT cross-entropy loss.
We call this design in-context because the learnable adapter queries and instruction tokens are placed in the same Transformer context, allowing each adapter query to attend directly to the instruction requirements when generating LoRA weights.
In this architecture, attention enables tighter interactions among different target-module positions, which may facilitate functional partitioning within LALMs~\cite{guo2026ahamask}.
This formulation also eliminates the need for a pretrained sentence-embedding model, substantially reducing computational cost relative to \citet{text2lora}.

\section{Data and Tasks}
Task diversity is crucial for compositional generalization.
In this study, we use SIFT-50M~\cite{pandey2025sift} for supervised fine-tuning.
It contains 50 million multilingual speech-text pairs with natural language instructions, covering diverse speech understanding tasks such as acoustic-level perception, multi-input comparison, semantic understanding and open-ended generation.
To narrow down the scope, we only train on the English subset and the closed-ended ``acoustic level'', ``comparison'' and ``content level'' task partitions.
This yields around 4.5 million samples for training.

\subsection{Compositional Tasks for Evaluation}
For evaluation, we first use EvalSIFT, the official evaluation set along with SIFT-50M, for in-domain task evaluation.
Similar to training, we only use the closed-ended ``acoustic level'', ``comparison'' and ``content level'' partitions in EvalSIFT, each of which contains 500 samples.

We then construct a compositional evaluation benchmark by composing diverse tasks from the atomic speech understanding tasks, to evaluate compositional generalization after training on SIFT. This benchmark includes:
\begin{itemize}
	\item \textbf{Atomic} speech understanding tasks: five basic single-input tasks, including automatic speech recognition (ASR), gender recognition (GR), speech emotion recognition (SER), language identification (LID), and spoken question answering (SQA). These functionalities are well represented either in the SIFT training set or in the pretrained LALM backbone.
	\item \textbf{Comparative} tasks: we compose the atomic tasks into comparative tasks by providing the model with two speech clips and asking it to compare on a given atomic acoustic dimension. 
    Note that the SIFT training set covers the two-input task schema itself, but not each comparative task instance.
	For example, ``Are the two speakers with the same gender?''
	\item \textbf{Selective execution} tasks: based on comparative tasks, we further ask the model to perform another acoustic functionality on the selected result. For example, ``identify the language of the speaker who speaks for a longer duration''.
	\item \textbf{Sequential execution} tasks: we ask the model to perform a single atomic functionality on multiple inputs, such as ``recognize the genders of the following three speakers''.
	\item \textbf{Multi-function} tasks: we ask the model to perform multiple atomic functionalities on a single input, such as ``recognize the language and gender of the speaker''.
\end{itemize}

Hence, the primitive components of these compositional tasks are well covered in the training set (or in the pretrained LALM backbone), from acoustic functionalities to task schemas.
The task instances are also simple and intuitive, such that an LALM with strong compositional generalization should easily master them without complex reasoning after being trained on SIFT.
We use standard test sets for the atomic tasks, and compose them to form the other compositional task schemas.
For ASR and gender-related tasks, we use LibriSpeech test-clean \cite{panayotov2015librispeech}; for emotion-related tasks, we use MELD test set \cite{poria2019meld}; for language-related tasks, we use MLS test set \cite{pratap20_interspeech}. We additionally use the CN-College-Listen SQA test set from AudioBench \cite{wang2025audiobench} as a selective execution task, because it requires both gender recognition and semantic understanding abilities.
A list of detailed task descriptions, source datasets, sample counts, etc., is provided in the appendix.

\begin{figure*}[t]
    \centering
    \includegraphics[width=\textwidth]{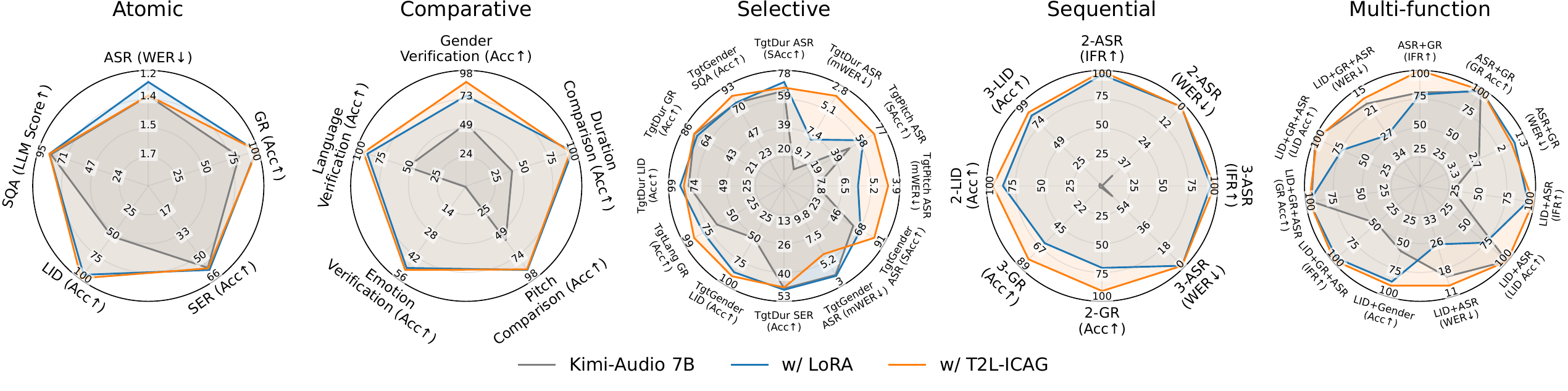}
    \caption{Radar plots for the five compositional task groups. Note that for the WER metric, the outer rings correspond to lower (better) values.
    The metrics are included in the parentheses.
    }
    \label{fig:radars-main}
\end{figure*}

\subsection{Train-Test Coverage}
Since the SIFT dataset itself is a diverse instruction-tuning dataset, we further audit how much of our compositional evaluation benchmark is covered by the training data.
We define a set of acoustic primitive functionalities, such as the perception of gender, pitch, content, etc.
We also define task schemas that are orthogonal to these acoustic primitives, such as single-input, comparison, multi-function tasks, etc.
Using these definitions, we tag the SIFT training samples with LLM assistance and count, for each evaluation task, how many training samples match the same primitive functionality set and task schema.
This gives the \textbf{primitive-level} coverage: the training data that uses the same acoustic functionalities and schema, but not necessarily the same exact task form.
We then conduct a stricter LLM-assisted audit for \textbf{exact-form} coverage, where the matched training samples must also share the concrete task form, including input structure, operation, output target, and explicit output schema when applicable.

\begin{table}[t]
    \centering
    \small
    \setlength{\tabcolsep}{8pt}
    \begin{tabular}{lcc}
    \toprule
    \textbf{Task group} & \textbf{Primitive I/S/O} & \textbf{Exact I/S/O} \\
    \midrule
    Atomic & 5 / 0 / 0 & 4 / 0 / 1 \\
    Comparative & 2 / 3 / 0 & 2 / 2 / 1 \\
    Selective & 1 / 1 / 7 & 0 / 0 / 9 \\
    Sequential & 4 / 2 / 0 & 1 / 1 / 4 \\
    Multi-function & 1 / 3 / 0 & 0 / 1 / 3 \\
    \midrule
    Total & 13 / 9 / 7 & 7 / 4 / 18 \\
    \bottomrule
    \end{tabular}
    \caption{Task distribution under overlap-based partitions: in-domain (I), sparse (S), and out-of-domain (O).}
    \label{tab:task-overlap-distribution}
\end{table}

For both coverage criteria, we label tasks with more than 1000 matched training samples as \textbf{in-domain}, tasks with 1--1000 matches as \textbf{sparse}, and tasks with zero matches as \textbf{OOD}.
Note that sparse tasks account for at most about 0.02\% of the training set and are therefore far from well represented.
Detailed primitive definitions and data audits are provided in the appendix.

Table \ref{tab:task-overlap-distribution} reveals a clear coverage gap: 13 of 29 tasks are in-domain and 7 are OOD at the primitive level, whereas only 7 are in-domain and 18 are OOD under the exact-form criterion.
All nine selective tasks are exact-form OOD, while most sequential and multi-function tasks overlap with SIFT only at the primitive level.
Thus, beyond familiar capabilities, the benchmark tests whether models can recombine known primitives into unseen or sparsely represented task forms.

\section{Experimental Results}
\subsection{Model, Training and Evaluation}
In the main experiments in this study, we use Kimi-Audio-7B~\cite{ding2025kimi} as the LALM backbone. We remove all audio-generation layers and focus only on the understanding part.
The hypernetwork uses rank 4, 4-layer ModernBERT~\cite{modernbert}-style Transformer encoder with hidden size 256, 8 attention heads, and rotary positional encoding. 
The resulting parameter count is about 30M.

We also set $r=4$ on all query and value projection layers for both static LoRA and T2L variants by default. For Kimi-Audio, this setup results in about 1.3M LoRA parameters.
Although our filtered SIFT training set contains a wide range of compositional tasks, it relies on a relatively small set of atomic audio capabilities. Many of these capabilities are already present in the pretrained Kimi-Audio backbone. Therefore, fine-tuning is mainly expected to activate underused capabilities and teach the model how to understand and combine task schemas, rather than to inject a large amount of new acoustic knowledge. 
We additionally evaluate a high-capacity static LoRA baseline with $r=80$, whose trainable parameter count is comparable to that of T2L-ICAG.
LoRA scaling $\alpha/r$ is fixed to 4 for both LoRA settings. 
Other training configurations are detailed in the appendix.

For ASR-related tasks, we ask the model to produce JSON-formatted outputs for easy parsing and word error rate (WER) calculation, so we also measure the instruction following rate (IFR) for JSON schema compliance.
For selective-execution ASR tasks, standard WER conflates target selection with transcription quality.
We therefore additionally report selection accuracy (SAcc) and minimum WER (mWER) to evaluate these two abilities separately.
Given a predicted transcript, we compute its WER against the reference transcript of each input utterance and infer that the model selected the utterance yielding the lower WER.
SAcc is the proportion of examples for which this inferred selection matches the task-specified target, whereas mWER is computed using the better-matching reference for each example, thereby measuring transcription quality independently of target selection.
For other tasks, we use LLM-as-a-judge~\cite{zheng2023judging} to avoid tedious answer normalization, where we use Qwen3-32B~\cite{yang2025qwen3} as the judge model. 
A consistency analysis of different judge models is provided in the appendix.

\begin{table*}[t]
    \centering
    \small
    \setlength{\tabcolsep}{4pt}
    \begin{tabular}{lccccccccc}
    \toprule
    \multirow{2}{*}{\textbf{Method}} & \multicolumn{3}{c}{\textbf{EvalSIFT (ACC\%$\uparrow$)}} & \multicolumn{6}{c}{\textbf{Our Compositional Evaluation (RPS\%$\uparrow$)}} \\
    \cmidrule(lr){2-4}\cmidrule(lr){5-10}
    & \textbf{Acoustic} & \textbf{Content} & \textbf{Comparison} & \textbf{Atomic} & \textbf{Comparative} & \textbf{Selective} & \textbf{Sequential} & \textbf{Multi-function} & \textbf{Avg.} \\
    \midrule
    Kimi-Audio 7B & 18.6 & 41.8 & 11.6 & 85.57 & 44.45 & 79.22 & 2.00 & 76.90 & 57.63 \\
    w/ LoRA ($r=4$) & 75.4 & 58.0 & 69.0 & \textbf{98.96} & 95.07 & 91.10 & 91.08 & 83.87 & 92.02 \\
    w/ LoRA ($r=80$) & \textbf{78.4} & \textbf{59.4} & 72.6 & 94.58 & 95.76 & 67.38 & 83.88 & 92.46 & 86.81 \\
    w/ T2L-emb & 77.2 & 59.2 & 71.0 & 97.43 & 90.59 & 87.62 & 92.87 & 86.42 & 90.99 \\
    w/ T2L-ICAG & 77.0 & 58.8 & \textbf{72.8} & 98.47 & \textbf{98.25} & \textbf{99.30} & \textbf{98.29} & \textbf{97.44} & \textbf{98.35} \\
    \bottomrule
    \end{tabular}
    \caption{Accuracy and RPS results across EvalSIFT categories and our compositional task classes. All T2L models use $r=4$.}
    \label{tab:main-results}
\end{table*}

As each task class contains multiple tasks with different evaluation metrics, we use relative performance score (RPS) as a fair aggregation, following \cite{peng2026survey,li2026isa}.
Given a metric and a group of candidate systems, let $M_i$ denote the metric score of candidate $i$.
For higher-is-better metrics like accuracy, we define $\mathrm{RPS}_i = M_i / \max_j (M_j)$; for lower-is-better metrics like WER, we define $\mathrm{RPS}_i = \min_j (M_j) / M_i$.
This score reflects the relative position of a candidate among all candidates, eliminating the effect of different metric scales.

\subsection{Overall Performance Across Task Schemas}

We first present the raw performance scores between the original Kimi-Audio, multi-task static LoRA trained on our filtered SIFT dataset, and T2L-ICAG.
We provide radar plots in Figure \ref{fig:radars-main} for the five compositional task classes.
We set the outer ring radius to the best value among the three displayed systems.
As shown in Figure \ref{fig:radars-main}, the original Kimi-Audio checkpoint performs well on atomic tasks, but degrades sharply when multiple speech inputs or task functions are introduced.
This is expected, as the original training set of this LALM likely does not contain these task schemas.
Multi-task LoRA improves performance across task classes.
T2L-ICAG stays on par with multi-task LoRA on atomic tasks, with a minor disadvantage on LibriSpeech test-clean WER (LoRA 1.27\% vs. T2L-ICAG 1.34\%) and a minor gain on LID accuracy (LoRA 95.20\% vs. T2L-ICAG 98.90\%).
On comparative tasks, T2L-ICAG achieves a major improvement over LoRA on gender verification (LoRA 76.32\% vs. T2L-ICAG 87.96\%).
On selective execution, sequential execution and multi-function tasks, T2L-ICAG achieves substantial gains in most cases,
e.g., 80.23\% vs. 62.60\% accuracy on the triple-input GR task and 99.70\% vs. 75.05\% LID accuracy on the LID+GR+ASR multi-function task.

To provide a quantitative aggregation of these scores, Table \ref{tab:main-results} presents the RPS scores for each task class.
This table reports both the in-domain EvalSIFT accuracies and our compositional evaluation RPS scores among the five candidates.
Under each task class, we compute the RPS for each task and take the average.

Table \ref{tab:main-results} shows that T2L-ICAG achieves the best average Kimi-Audio performance across the compositional task classes.
On in-domain tasks such as EvalSIFT evaluations and atomic tasks where the training set has sufficient exact-task coverage, T2L-ICAG achieves similar performance with multi-task LoRA and sometimes mild improvements.
On more novel task forms, the advantage of T2L-ICAG is more significant, especially on sequential (+7.2 RPS points) and multi-function tasks (+13.6 RPS points) relative to rank-4 LoRA.
The parameter-matched rank-80 LoRA variant obtains even lower RPS scores compared to $r=4$, despite noticeable gains on in-domain EvalSIFT evaluations.

These results demonstrate that T2L-ICAG can effectively generate task-specific LoRA adapters from instructions in multiple task forms. 
Using atomic acoustic functionalities as building blocks, the hypernetwork can compose unseen tasks from existing functionalities and transfer learned knowledge to novel task compositions better than static LoRA adapters.
Also, the generalization gains of T2L-ICAG arise from the adaptation mechanism rather than merely from a larger number of trainable parameters.

\subsection{Generalization Beyond Training Coverage}
To connect performance with training coverage, Table \ref{tab:kimi-audit-rps} further reports Kimi-Audio results under the train-test overlap partitions defined in Table \ref{tab:task-overlap-distribution}.
We compute RPS within each partition in the same way as in Table \ref{tab:main-results}.
These results show that the advantage of T2L-ICAG is not limited to tasks with high train-test overlap.
Under primitive-level grouping, T2L-ICAG achieves consistently high RPS across in-domain, sparse and OOD partitions, indicating that it benefits not only from familiar primitive combinations but also from tasks with no matched primitive-level training samples.
Under the stricter exact-form partitioning, T2L-ICAG still achieves the best score in all three coverage regimes.
In particular, it obtains a 9-point gain on exact-form OOD tasks over rank-4 LoRA, showing that its advantage is not explained only by repeated exposure to the same task forms during training.
The parameter-matched $r=80$ LoRA baseline also remains below T2L-ICAG across all six coverage partitions.
This suggests that conditioning adapter generation on the full instruction structure helps the model transfer learned acoustic primitives to novel task forms, rather than merely memorizing frequent training instructions.

\begin{table}[t]
\centering
\small
\setlength{\tabcolsep}{4pt}
\begin{tabular}{lccc}
\toprule
\textbf{Method} & \textbf{In-domain} & \textbf{Sparse} & \textbf{OOD} \\
\midrule
\multicolumn{4}{l}{\textbf{Primitive-level}} \\
Kimi-Audio 7B & 52.42 & 58.05 & 74.44 \\
w/ LoRA ($r=4$) & 93.20 & 87.57 & 90.12 \\
w/ LoRA ($r=80$) & 86.59 & 87.84 & 77.78 \\
w/ T2L-emb & 92.50 & 88.39 & 87.47 \\
w/ T2L-ICAG & \textbf{97.78} & \textbf{98.53} & \textbf{99.09} \\
\midrule
\multicolumn{4}{l}{\textbf{Exact-form}} \\
Kimi-Audio 7B & 64.76 & 28.16 & 61.44 \\
w/ LoRA ($r=4$) & 93.46 & 96.13 & 88.98 \\
w/ LoRA ($r=80$) & 93.88 & 91.09 & 82.40 \\
w/ T2L-emb & 92.36 & 71.83 & 92.17 \\
w/ T2L-ICAG & \textbf{98.21} & \textbf{99.52} & \textbf{98.14} \\
\bottomrule
\end{tabular}
\caption{RPS scores under primitive-level and exact-form train-test overlap partitions.}
\label{tab:kimi-audit-rps}
\end{table}

\subsection{Compositional Transfer of Atomic Functionalities}

Beyond grouping tasks by their schema or train-test overlap, we further examine whether a model can preserve an atomic capability when it is embedded into different compositional task forms.
For example, if the model's atomic gender-recognition (GR) accuracy is $p$, then a simple independent estimate is $p^2$ accuracy for recognizing the genders of two speakers (2-GR) and $p^3$ for three speakers (3-GR).
For a gender-verification task, where the model needs to decide whether two speakers have the same gender, the corresponding expectation becomes $p^2 + (1-p)^2$ under balanced labels.
These quantities are not strict upper bounds in all settings, since the exact value depends on the task construction, label distribution and possible shortcut effects.
However, they provide a useful reference for asking whether a LALM can carry an atomic recognition ability into a compositional instruction without substantial degradation.

\begin{figure}[t]
    \centering
    \includegraphics[width=\linewidth]{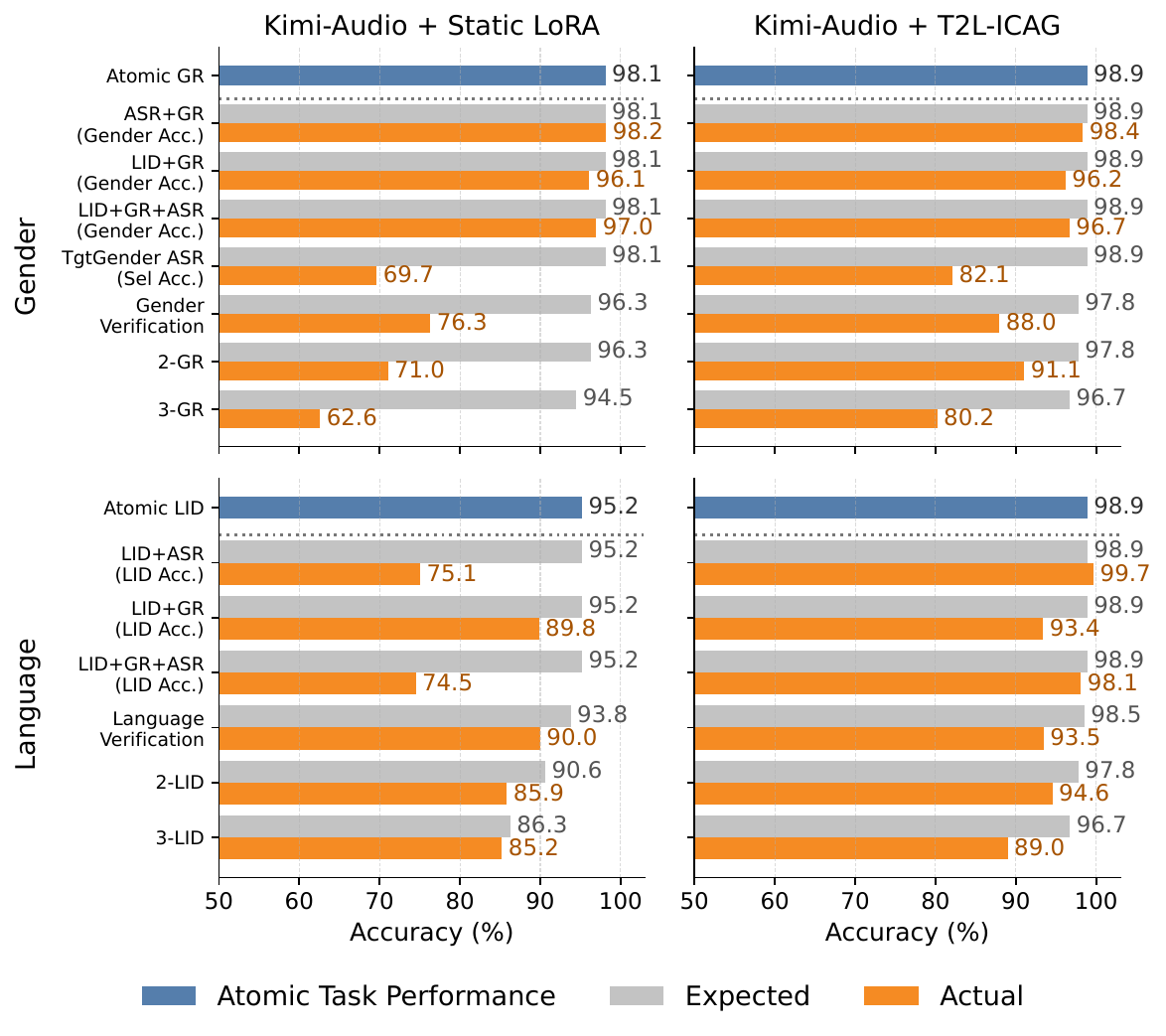}
    \caption{Gender- and language-related compositional performance compared with the expected performance implied by atomic recognition ability. The ``expected'' bars are calculated from atomic performance.}
    \label{fig:gender-language-expected-accuracy}
\end{figure}

Figure \ref{fig:gender-language-expected-accuracy} compares static LoRA and T2L-ICAG on gender- and language-related task families.
While the atomic recognition accuracies are comparable between static LoRA and T2L-ICAG, T2L-ICAG is consistently closer to the expected compositional performance on structurally harder variants.
The gap is especially visible on multi-input tasks such as 2-GR, 3-GR, 2-LID and 3-LID, as well as selection-style gender tasks, where static LoRA suffers a much larger drop from the atomic reference.
This suggests that T2L-ICAG improves not merely by strengthening the underlying acoustic classifier, but by better conditioning the adapter on the compositional structure expressed in the instruction.
Nevertheless, the remaining gaps between actual and expected performance show that compositional generalization is still far from solved, leaving more robust composition of atomic audio capabilities as important future work.

\subsection{Ablation Studies}

\subsubsection{Varying LoRA Ranks}

\begin{figure}[!t]
    \centering
    \includegraphics[width=\linewidth]{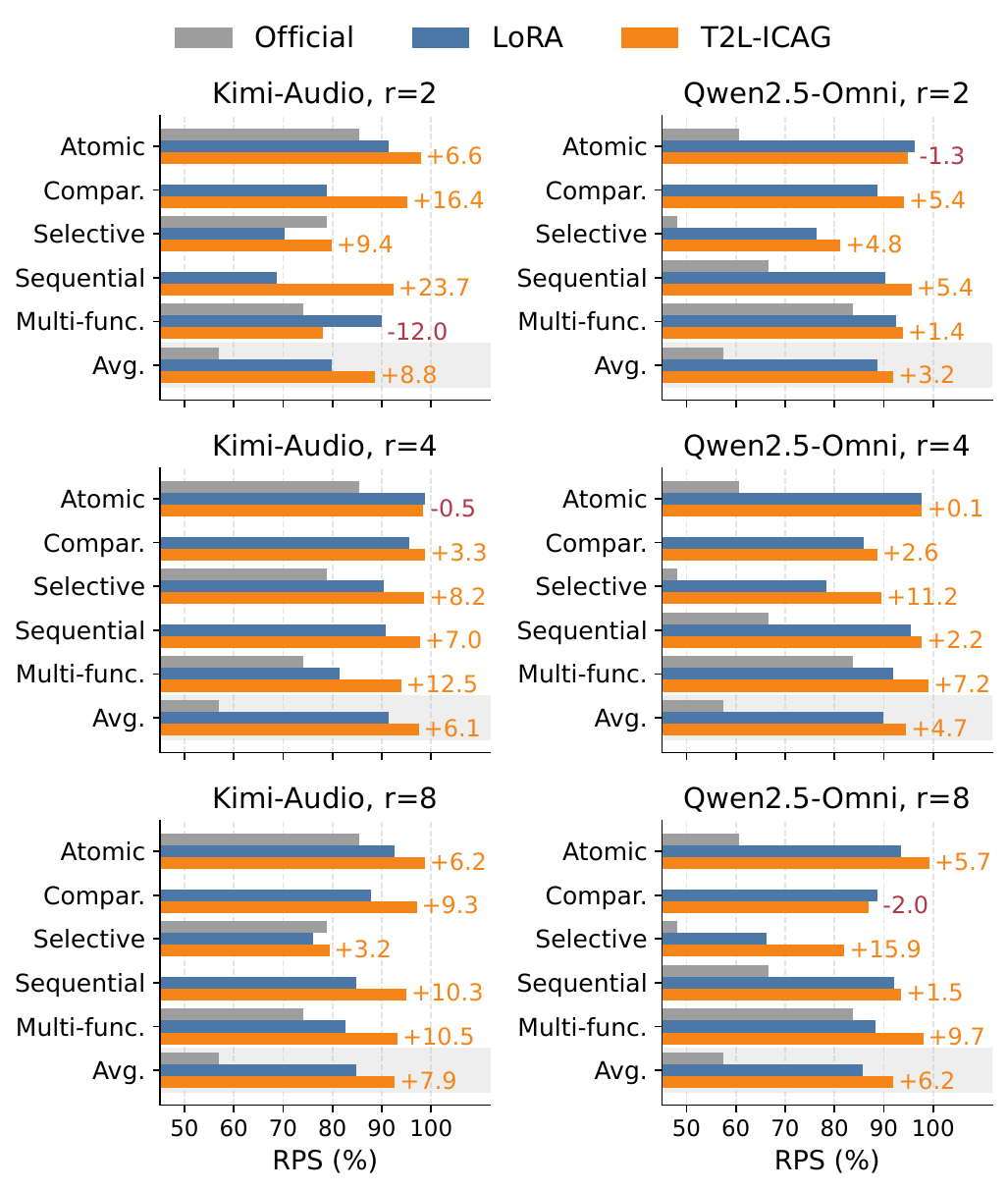}
    \caption{Rank ablation results across compositional task groups. Each panel corresponds to one backbone model and one LoRA rank. The shaded region marks the average over the five task groups. The numbers after bars denote the gain of T2L-ICAG over LoRA.}
    \label{fig:rank-ablation}
\end{figure}

We also evaluate T2L-ICAG under different LoRA ranks.
Figure \ref{fig:rank-ablation} shows the RPS results with ranks 2, 4 and 8 on both Kimi-Audio and Qwen2.5-Omni~\cite{xu2025qwen25omnitechnicalreport} thinker.
We compute RPS scores across all ranks within each task class and model backbone.
Each panel then compares the original checkpoint, static multi-task LoRA and T2L-ICAG under one backbone-rank setting, while the shaded row reports the average over the five task groups.

Several patterns can be observed.
First, static multi-task LoRA indeed improves over the original checkpoints after training on SIFT, a versatile multi-task dataset.
Second, T2L-ICAG further improves over static LoRA in the average score for every tested rank and backbone.
The average gains are especially clear for Kimi-Audio, where T2L-ICAG improves over LoRA by 8.8\%, 6.1\% and 7.9\% RPS points at ranks 2, 4 and 8, respectively.
For Qwen2.5-Omni, the corresponding gains are 3.2\%, 4.7\% and 6.2\% RPS points.
Note that these average scores aggregate more than 30 task metrics, indicating that the gains extend beyond a few isolated tasks.
The improvements are usually larger on more compositionally demanding groups, such as selective execution, sequential execution and multi-function tasks, which contribute the largest gains over LoRA.
There are still group-level exceptions, such as the multi-function group for Kimi-Audio at rank 2 and the comparative group for Qwen2.5-Omni at rank 8, but these do not change the overall average trend.
Therefore, across backbone models and ranks, T2L-ICAG generally improves over static LoRA on compositional task groups, indicating that the gain does not rely on a particular adapter rank.

We further show in the Appendix that increasing the LoRA rank beyond 8 does not improve generalization on this compositional benchmark.
Importantly, increased capacity does not lead to better compositional generalization ability for static LoRA adapters.

\subsubsection{Varying Hypernetwork Architectures}
We further compare T2L-ICAG with the global embedding-based T2L-emb variant in Table \ref{tab:main-results} and Table \ref{tab:kimi-audit-rps}.
This variant follows the original T2L design, where the instruction is first compressed into a single sentence embedding, and the hypernetwork generates LoRA weights from this global vector.

The embedding-based variant performs competitively on in-domain EvalSIFT tasks and sometimes slightly outperforms T2L-ICAG there.
However, its compositional generalization is weaker.
On Kimi-Audio, T2L-emb obtains an average compositional RPS of 90.99, below both rank-4 static LoRA (92.02) and T2L-ICAG (98.35).
The gap is especially clear on structurally complex groups such as selective execution, sequential execution and multi-function tasks.
In Table \ref{tab:kimi-audit-rps}, T2L-emb also exhibits weaker generalization in sparsely covered and OOD scenarios.
These results suggest that a single global instruction embedding is insufficient for modeling fine-grained task composition in LALMs, given the training set.
The ICAG design preserves token-level task structure and provides more effective conditioning for generating task-specific adapters.

\section{Conclusion}

In this work, we study compositional generalization in LALMs by testing whether familiar atomic acoustic functionalities can be recombined under novel task compositions.
To improve such transfer, we propose T2L-ICAG, which generates LoRA adapters from token-level task information through the ICAG hypernetwork.
Across multiple compositional task schemas, T2L-ICAG consistently outperforms static LoRA and previous T2L-emb, with clear gains on structurally demanding tasks.
Our overlap and atomic-functionality analyses further suggest that these gains come from better composition of known primitives rather than simple memorization or more parameters.

Nevertheless, robust generalization to unseen task compositions remains an open challenge.
Additional experiments, especially a singular-value analysis of trained and generated LoRA weights, are provided in the appendix.

\twocolumn[%
  \centering
  \vspace{0.2in}%
  {\LARGE\bf Appendix\par}%
  \vspace{0.15in}%
]
In this appendix, we provide details on our dataset construction, evaluation protocol, detailed experiment setups, and more experimental results on LoRA ranks, translation tasks, mismatched instructions, state-of-the-art LALMs, singular value analysis, and LLM judge consistency.

\section{Data and Task Details}

\subsection{Task Definitions and Evaluation Metrics}
Table~\ref{tab:data-task-details} and Table~\ref{tab:data-task-details-continued} show the details of task classes, source datasets, task requirements, and evaluation protocols.
Specifically, the atomic tasks are standard single speech-input tasks using widely-used test sets.
For the SQA task here, we use the SLUE-phase2-SQA5~\cite{shon2023slue} test set, following AudioBench~\cite{wang2025audiobench}.

The comparative tasks extend atomic ones into double speech inputs and ask the LALM to compare the two speakers along an acoustic attribute.
We also introduce pitch and duration comparison tasks in this class.
We compose those comparative tasks using the same test sets as the atomic tasks.
For pitch comparison tasks, we first use the YIN~\cite{de2002yin} algorithm to extract the pitch contours of the speech in LibriSpeech~\cite{panayotov2015librispeech} test-clean, and create speaker pairs whose average pitches differ by at least 4 semitones. 
For duration comparison tasks, we form pairs using LibriSpeech test-clean where the duration of the two speakers differ by at least 3 seconds.
For emotion verification, we construct evaluation pairs using the MELD~\cite{poria2019meld} test set.
Positive pairs are formed from utterances sharing the same emotion label. 
To reduce ambiguity among semantically similar emotions, negative pairs are sampled only from emotion categories that are sufficiently separated in the affective space (e.g., angry and sad), while highly similar emotion pairs (e.g., joy and surprise) are excluded.

The selective execution tasks extend comparative tasks from a single task to two tasks. These tasks ask the LALM to first determine the target speaker based on the first acoustic attribute, and then perform the second task on that. 
There are plenty of task combinations for this class, but we only keep the meaningful ones and exclude artificial ones\footnote{For example, asking the LALM if the female speaker has a higher average pitch than the male. As another example, evaluating ASR and emotion together on MELD utterances might lead to high WER scores because MELD samples are usually short and colloquial. }. 
One special case is the target gender SQA task, where we use the CN-College-Listen partition from AudioBench~\cite{wang2025audiobench}. 
This test set requires the LALM to identify the activities and content from the target-gender speaker in a two-person dialogue, and answer a question about them\footnote{For example, ``What does the female speaker say about the book?''}.
This is different from standard SQA tasks like SLUE, where the speech is from a single speaker.

For target-$X$ ASR tasks (target-gender / duration / pitch ASR), the model receives two utterances and must transcribe only the speaker selected by attribute $X$.
A naive target-conditioned WER conflates two failure modes: selecting the wrong speaker versus poorly transcribing the correct one.
We therefore report two complementary metrics that separate these abilities.
Given a prediction $\hat{y}$ and the two references $(y_1, y_2)$, we compute $\mathrm{WER}(\hat{y}, y_1)$ and $\mathrm{WER}(\hat{y}, y_2)$, and treat the lower-WER reference as the model's \emph{inferred} selection.
\textbf{Selection accuracy (SAcc)} is the fraction of examples for which this inferred selection matches the task-specified target speaker.
\textbf{Minimum WER (mWER)} is the corpus-level WER computed against the better-matching reference for each example, i.e., an oracle-matched WER that measures transcription quality independently of target selection.\footnote{This is related to, but distinct from, the concatenated minimum-permutation WER (cpWER)~\cite{watanabe2020chime} commonly used in speaker diarization with ASR, which evaluates multi-speaker transcripts by choosing the speaker ordering that yields the lower WER. In contrast, our setting produces a single predicted transcript under an attribute-conditioned selection criterion, so we do not call the metric cpWER.}
Intuitively, high SAcc with low mWER indicates both correct selection and faithful transcription, whereas low SAcc with low mWER suggests that the model can recognize speech but fails to follow the selection criterion.

We then explore multi-input and multi-task scenarios. 
In these cases, we consider the extension and composition of three atomic functionalities: ASR, gender recognition, and language identification.
For example, the 2ASR JSON task asks the LALM to transcribe two speakers into JSON format\footnote{With predefined keys, such as \texttt{\{"Speaker1": "Content", "Speaker2": "Content"\}}.}, and the 2-GR task asks the LALM to recognize the genders of two speakers.
For all ASR-related tasks in multi-input or multi-task scenarios, we ask the model to answer in JSON format to avoid wrong WER scores. This introduce the instruction following rate (IFR) metric to evaluate the JSON format compliance.
We then calculate ASR WER in different cases:
\begin{itemize}
    \item In multi-input single-task scenarios, we introduce a \textbf{lenient} WER metric so that we can tolerate JSON formatting errors. That is, we normalize all the JSON special tokens and keys from the answer, no matter if the JSON is well-formatted or not, and then calculate the WER against the reference.
    \item In single-input multi-task scenarios, we have to calcualte the WER within instruction-following samples to avoid the interference of the other tasks.
\end{itemize}
In multi-task cases, when the output is a JSON format, we can evaluate the accuracy of language and gender tasks by simple string matching; otherwise, we use the LLM-as-a-judge to avoid normalization.

In this way, we construct a multi-faceted compositional task space from only 7 atomic acoustic understanding functionalities: content recognition, gender recognition, emotion recognition, pitch perception, duration perception, language identification, and semantic understanding.

\subsection{Task Primitives and Data Audits}
Since SIFT is released without detailed task-level annotations, and it only provides instructions in natural language, we could only use customized audits to analyze the train-test coverage.
To analyze the functional composition of an audio understanding task\footnote{More specifically, we mainly consider \textbf{speech} understanding tasks here since the SIFT training data focuses on speech.}, we define the following two closed sets:
\begin{itemize}
\item \textbf{Acoustic Primitives}, which is the basic acoustic understanding ability. These are: speech recognition, phoneme recognition, accent classification, language identification, emotion recognition, gender recognition, average pitch perception, pitch variation perception, translation, age recognition, sound quality and environment perception, duration perception, speaking rate perception, volume perception, general semantic understanding\footnote{For example, a speech question answering task where the model needs to answer based on the spoken content. The model does not need to transcribe every spoken word as-is, but should comprehend the spoken semantics.}, overlapped speech separation, event detection.
\item \textbf{Task Schemas}, which is the structure of an understanding task, independent of the acoustic primitives\footnote{There are also few exceptions, e.g. attributes like pitch and sound quality mostly appear in comparison instead of single-input tasks.}. These are: single-input, multi-input, single-task, multi-task, comparison. 
\end{itemize}
Although these are not enough to cover the whole speech and audio understanding space, they are sufficient enough to cover the subspace spanned by SIFT. 
We then ask an LLM, Qwen3-32B in our case, to annotate each training sample by choosing one or more items from both the acoustic primitives and task schemas.
For example:
\begin{itemize}
\item ``What is the accent of the person?'' $\to$ {[``single-input'', ``single-task'', ``accent classification'']}.
\item ``Which audio clip has a larger distortion level?'' $\to$ [``multi-input'', ``comparison'', ``sound quality and environment perception''].
\item ``How would you describe the tone and volume of the speaker in the audio?'' $\to$ [``single-input'', ``multi-task'', ``pitch variation perception'', ``volume perception''].
\end{itemize}

Then, for each evaluation task, we perform data audit to count the number of training samples that cover this task in two levels:
\begin{itemize}
	\item \textbf{Primitive-level}, where the sets of acoustic primitives and task schemas are matched;
	\item \textbf{Exact-form}, where the exact ordering of applying these functionalities is matched. Note that audits in this level could only be done via LLM agents by first setting up some heuristic string templates, and then refining the retrieval results with several rounds.
\end{itemize}

\begin{figure}[t]
    \centering
    \includegraphics[
        width=\linewidth
    ]{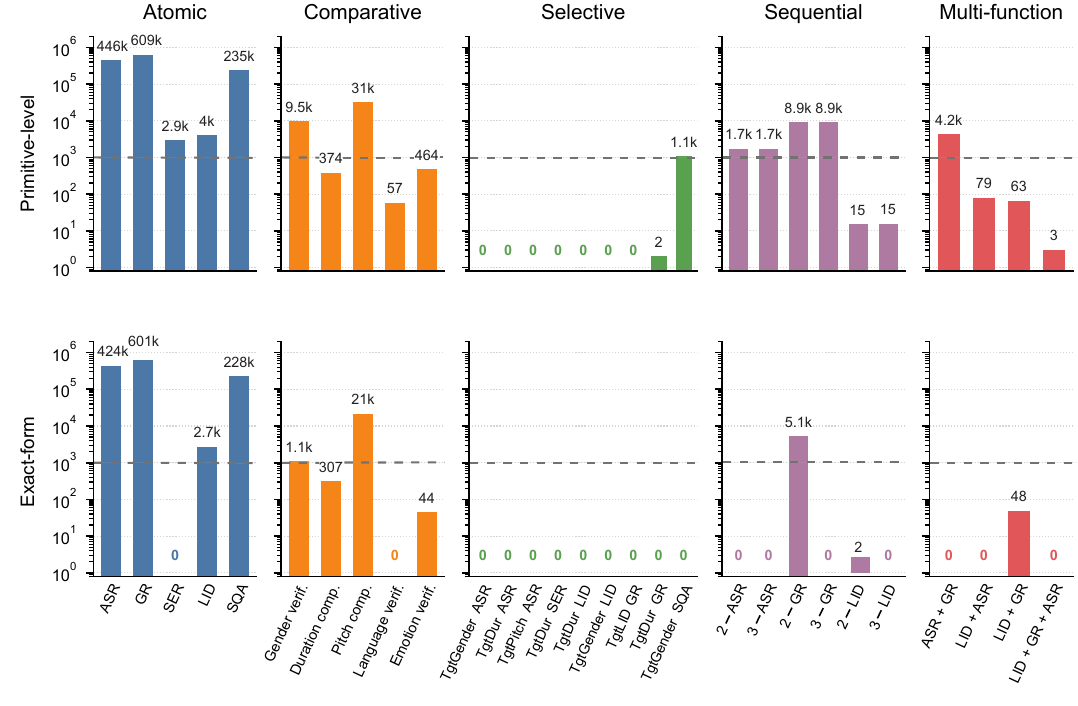}
    \caption{
        Training-set coverage of the 29 evaluation tasks, grouped
        by the five task groups.
        Bar heights show training-sample multiplicity on a shared
        logarithmic scale. Tasks with no matching training samples
        are marked with ``0''.
        Dashed lines denote the 1000-sample threshold, which we use to distinguish sparse tasks and in-domain tasks.
    }
    \label{fig:train-test-overlap-coverage}
\end{figure}

Figure \ref{fig:train-test-overlap-coverage} shows the exact number of matches for both the primitive-level and exact-form audits.

\begin{table*}[!htbp]
\centering
\renewcommand{\arraystretch}{1.12}
\setlength{\tabcolsep}{3pt}
\begin{tabular}{L{0.13\textwidth} L{0.20\textwidth} L{0.13\textwidth} L{0.075\textwidth} L{0.25\textwidth} L{0.10\textwidth}}
\toprule
\textbf{Task class} & \textbf{Task name} & \textbf{Source dataset} & \textbf{Sample num.} & \textbf{Requirements} & \textbf{Evaluation} \\

\midrule
\multirow{5}{0.12\textwidth}{\taskclass{Atomic task}{(single speech input, single task)}}
& \graycell{ASR (Automatic speech recognition)} & \graycell{LS test-clean} & \graycell{2620} & \graycell{Transcribe the spoken content into text.} & \graycell{WER} \\
& GR (Gender recognition) & LS test-clean & 2620 & Recognize speaker gender (male, female). & LLM-judged Acc.\\
& \graycell{SER (Speech emotion recognition)} & \graycell{MELD test} & \graycell{2610} & \graycell{Recognize speaker emotion (anger, disgust, fear, joy, neutral, sadness, surprise).} & \graycell{LLM-judged Acc.} \\
& LID (Language identification)& \stackcell{LS test-clean,\\MLS test} & 2000 & Identify speaker language (English, French, Italian, German, Spanish). & LLM-judged Acc.\\
& \graycell{SQA (Spoken question answering)} & \graycell{SLUE-p2-SQA5} & \graycell{408} & \graycell{Answer the text question based on the speech content.} & \graycell{LLM-judge Score} \\
\midrule
\multirow{5}{0.12\textwidth}{\taskclass{Comparative task}{(double speech inputs, single task)}}
& Gender Verification & LS test-clean & 5240 & Verify whether the two speakers have the same gender. & LLM-judged Acc.\\
& \graycell{Duration Comparison} & \graycell{LS test-clean} & \graycell{2094} & \graycell{Identify which speaker speaks longer.} & \graycell{LLM-judged Acc.} \\
& Pitch Comparison & LS test-clean & 2372 & Identify which speaker has a higher average pitch. & LLM-judged Acc.\\
& \graycell{Emotion Verification} & \graycell{MELD test} & \graycell{2600} & \graycell{Verify whether the two speakers express the same emotion.} & \graycell{LLM-judged Acc.} \\
& Language Verification & \stackcell{LS test-clean,\\MLS test} & 4350 & Verify whether the two speakers speak the same language. & LLM-judged Acc. \\
\midrule
\multirow{9}{0.12\textwidth}{\taskclass{Selective execution}{(double speech inputs, two tasks)}}
& \graycell{Target Gender LID} & \graycell{\stackcell{LS test-clean,\\MLS test}} & \graycell{2000} & \graycell{Identify the target-gender speaker's language.} & \graycell{LLM-judged Acc.} \\
& Target Language GR & \stackcell{LS test-clean,\\MLS test} & 1589 & Identify the target-language speaker's gender. & LLM-judged Acc. \\
& \graycell{Target Gender ASR} & \graycell{LS test-clean} & \graycell{2620} & \graycell{Transcribe the target-gender speaker.} & \graycell{SAcc, mWER} \\
& Target Duration GR & LS test-clean & 2000 & Recognize the longer-duration speaker's gender. & LLM-judged Acc.\\
& \graycell{Target Duration ASR} & \graycell{LS test-clean} & \graycell{2094} & \graycell{Transcribe the longer-duration speaker.} & \graycell{SAcc, mWER} \\
& Target Duration SER & MELD test & 2610 & Recognize the longer-duration speaker's emotion. & LLM-judge \\
& \graycell{Target Duration LID} & \graycell{\stackcell{LS test-clean,\\MLS test}} & \graycell{2000} & \graycell{Identify the longer-duration speaker's language.} & \graycell{LLM-judge} \\
& Target Pitch ASR & LS test-clean & 2372 & Transcribe the higher-pitch speaker. & SAcc, mWER \\
& \graycell{Target Gender SQA} & \graycell{\stackcell{AudioBench\\CN-College-\\Listen}} & \graycell{2271} & \graycell{Find the target-gender speaker and answer multiple-choice questions.} & \graycell{LLM-judged Acc.} \\
\bottomrule
\end{tabular}
\caption{Details of task classes, source datasets, instructions, and evaluation protocols. ``LS'' denotes LibriSpeech.}
\label{tab:data-task-details}
\end{table*}

\begin{table*}[!htbp]
\centering
\renewcommand{\arraystretch}{1.12}
\setlength{\tabcolsep}{3pt}
\begin{tabular}{L{0.13\textwidth} L{0.14\textwidth} L{0.14\textwidth} L{0.075\textwidth} L{0.26\textwidth} L{0.15\textwidth}}
\toprule
\textbf{Task class} & \textbf{Task name} & \textbf{Source dataset} & \textbf{Sample num.} & \textbf{Requirements} & \textbf{Evaluation} \\
\midrule
\multirow{6}{0.12\textwidth}{\taskclass{Sequential execution task}{(multiple speech inputs, single task)}}
& 2-ASR JSON & LS test-clean & 2620 & Transcribe two speakers into JSON format. & IFR, WER \\
& \graycell{3-ASR JSON} & \graycell{LS test-clean} & \graycell{2620} & \graycell{Transcribe three speakers into JSON format.} & \graycell{IFR, WER} \\
& 2-GR & LS test-clean & 2620 & Recognize the genders of two speakers. & LLM-judged Acc. \\
& \graycell{3-GR} & \graycell{LS test-clean} & \graycell{2620} & \graycell{Recognize the genders of three speakers.} & \graycell{LLM-judged Acc.} \\
& 2-LID & \stackcell{LS test-clean,\\MLS test} & 2100 & Identify the languages of two speakers. & LLM-judged Acc. \\
& \graycell{3-LID} & \graycell{\stackcell{LS test-clean,\\MLS test}} & \graycell{2100} & \graycell{Identify the languages of three speakers.} & \graycell{LLM-judged Acc.} \\
\midrule
\multirow{4}{0.12\textwidth}{\taskclass{Multi-functionality task}{(single speech input, multiple tasks)}}
& \stackcell{ASR + GR JSON} & LS test-clean & 2620 & Transcribe speech and recognize gender in JSON format. & \stackcell{IFR, WER,\\Gender Acc.} \\
& \graycell{\stackcell{LID + GR}} & \graycell{\stackcell{LS test-clean,\\MLS test}} & \graycell{2000} & \graycell{Identify language and recognize gender.} & \graycell{LLM-judged Acc.} \\
& \stackcell{LID + ASR JSON} & \stackcell{LS test-clean,\\MLS test} & 2000 & Identify language and transcribe speech in JSON format. & \stackcell{IFR, WER ,\\Language Acc.} \\
& \graycell{\stackcell{LID + GR + ASR JSON}} & \graycell{\stackcell{LS test-clean,\\MLS test}} & \graycell{2000} & \graycell{Identify language, recognize gender, and transcribe speech in JSON format.} & \graycell{\stackcell{IFR, WER, \\Language Acc.,\\Gender Acc.}} \\
\bottomrule
\end{tabular}
\caption{(Continued) Details of task classes, source datasets, instructions, and evaluation protocols. ``LS'' denotes LibriSpeech.}
\label{tab:data-task-details-continued}
\end{table*}

\section{Model and Training Details}

\subsection{Model Details}
Table~\ref{tab:backbone-configurations} summarizes the configurations of Kimi-Audio and the corresponding ICAG hypernetwork in our experiments. Note that Kimi-Audio uses grouped query attention. We discard the 6 audio generation layers since they are irrelevant to understanding. For Qwen2.5-Omni, we only experiment on the thinker part, which has the same configurations with Kimi-Audio since they share a Qwen2-7B implementation backbone.

\begin{table}[t]
\centering

\setlength{\tabcolsep}{4pt}
\begin{tabular}{lcc}
\toprule
\begin{tabular}[c]{@{}c@{}}\textbf{Configuration}\end{tabular} &
\begin{tabular}[c]{@{}c@{}}\textbf{Kimi-Audio 7B}\end{tabular} &
\begin{tabular}[c]{@{}c@{}}\textbf{ICAG}\\\textbf{Hypernetwork}\end{tabular} \\
\midrule
Num. Layers & 28 & 4 \\
Hidden Dim. & 3584 & 256 \\
Head Dim. & 128 & 32 \\
Num. KV Heads & 4 & 8 \\
Num. Q Heads & 28 & 8 \\
\bottomrule
\end{tabular}
\caption{Configurations of the LALM backbone and ICAG hypernetwork. For Qwen2.5-Omni Thinker, the configurations are exactly the same.}
\label{tab:backbone-configurations}

\end{table}

\subsection{Training Details}

\begin{figure}[t]
\centering
\includegraphics[width=\linewidth]{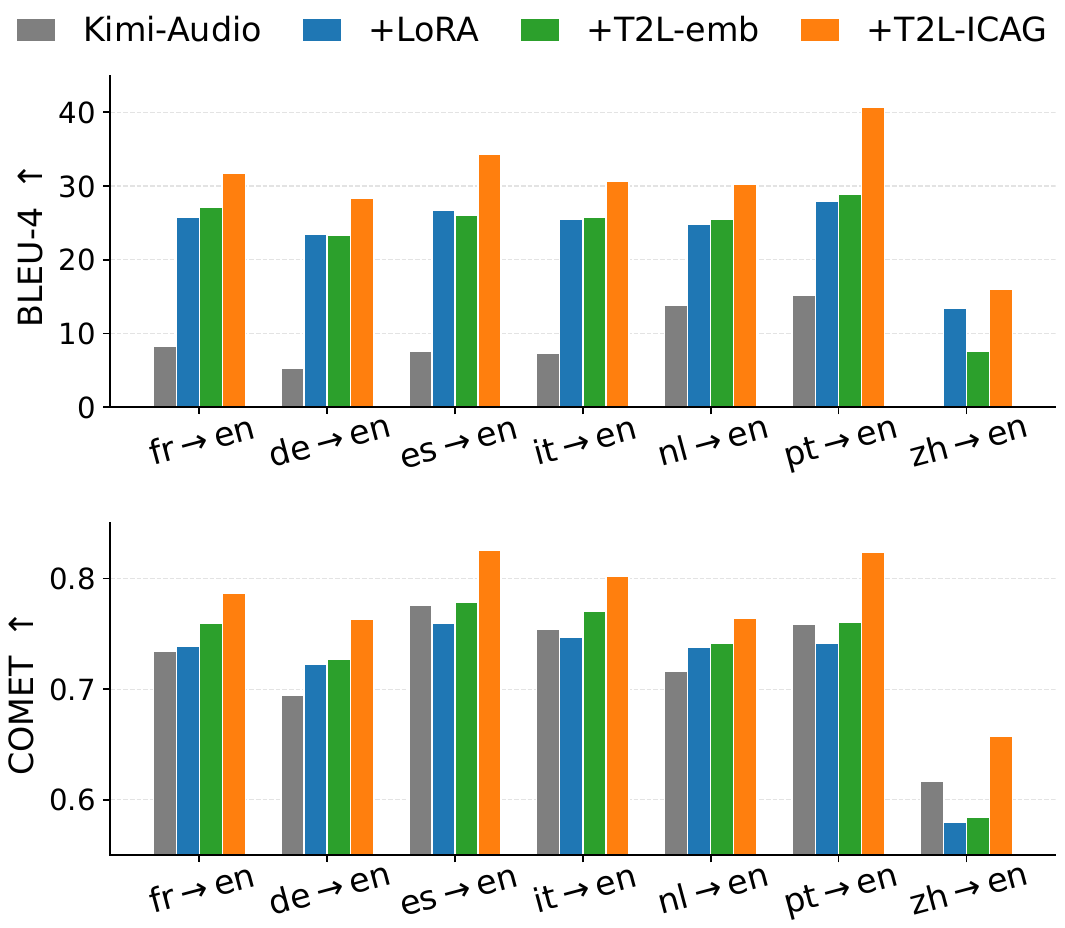}
\caption{BLEU-4 and COMET results on CoVoST2 X2En speech-to-text translation from seven source languages: French (fr), German (de), Spanish (es), Italian (it), Dutch (nl), Portuguese (pt), and Mandarin (zh). The Kimi-Audio bar is invisible for zh$\rightarrow$en under BLEU-4 because its score is 0.}
\label{fig:covost2-translation}
\end{figure}

We set the LoRA dropout rate to 5\%.
We train for one epoch with global batch size 16, AdamW~\cite{adamw} optimizer, learning rate $5\times10^{-5}$, weight decay 0.01, cosine learning-rate scheduler with minimum learning rate $10^{-6}$, warmup ratio 1\%, and gradient clipping at 1.0. 
Training hardware is 4 NVIDIA A800 GPUs with 80GB VRAM. Training runs use a fixed random seed of 42 in PyTorch and distributed data shuffling.
DeepSpeed ZeRO-2 is enabled.

\subsection{Inference Settings}
We use greedy decoding for deterministic evaluation; max number of new tokens is set to 300, which is sufficiently large for our evaluation tasks.

\section{More Experiments}
\subsection{Larger LoRA Ranks Are Not Enough}
\begin{figure}[t]
    \centering
    \includegraphics[width=\linewidth]{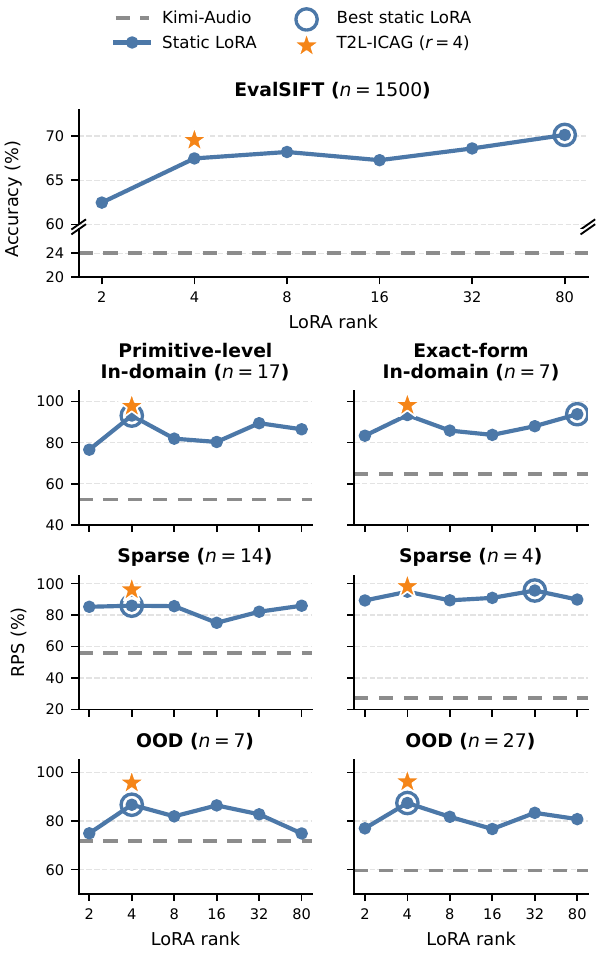}
    \caption{Effect of static LoRA rank on EvalSIFT and compositional generalization under different train--test overlap partitions. The blue curves show static LoRA results across ranks, the gray dashed lines denote the original Kimi-Audio backbone, and the orange stars denote T2L-ICAG with $r=4$. A blue hollow circle marks the best static LoRA rank in each panel. The numbers $n$ indicate the number of evaluation metric units in each partition. The EvalSIFT panel reports raw accuracy, whereas the remaining panels report RPS.}
    \label{fig:lora-rank-sweep-overlap}
\end{figure}

We further vary the rank of static LoRA over $r\in\{2,4,8,16,32,80\}$ to examine whether increasing adapter capacity is sufficient to improve compositional generalization.
Figure~\ref{fig:lora-rank-sweep-overlap} reports the overall EvalSIFT accuracy and the RPS under the primitive-level and exact-form train--test overlap partitions.
For consistent comparison, the RPS values are jointly computed over Kimi-Audio, all six static LoRA ranks, and T2L-ICAG with $r=4$.

On the fully in-domain EvalSIFT benchmark, larger ranks generally improve static LoRA performance: the overall accuracy increases from 62.47\% at $r=2$ to 70.13\% at $r=80$, with only small intermediate fluctuations.
A similar high-rank benefit appears in the exact-form in-domain partition, where $r=80$ is the strongest static LoRA configuration.
This suggests that additional adapter capacity can help fit task forms that are already well represented in the training distribution, although the primitive-level in-domain results remain non-monotonic.

The trend changes under distribution shift.
For both primitive-level and exact-form OOD tasks, static LoRA achieves its best result at $r=4$; increasing the rank beyond 4 does not yield consistent gains and can substantially reduce RPS.
Moreover, rank-4 T2L-ICAG outperforms the best static LoRA configuration in all six overlap partitions.
These results show that simply increasing static LoRA capacity is insufficient for compositional generalization, while instruction-conditioned adapter generation provides gains that cannot be explained by rank alone.

\subsection{Translation Tasks}
Translation capability is also a core functionality of LALMs.
We observe that the Kimi-Audio backbone~\cite{ding2025kimi} often refuses to perform speech-to-text translation (S2TT), instead returning the source-language transcription.
Interestingly, although our SIFT training set~\cite{pandey2025sift} does not include translation tasks, both multi-task LoRA~\cite{hu2022lora} and T2L-based adaptation~\cite{text2lora} show non-trivial generalization to S2TT.
To evaluate this transfer ability, we use seven X-to-English translation directions from the CoVoST2 X2En benchmark~\cite{wang2021covost2}.
CoVoST2 covers a broad set of source languages, including several low-resource languages; here we focus on seven widely used non-English languages to avoid conflating adapter generalization with extremely low-resource translation behavior.

As shown in Figure~\ref{fig:covost2-translation}, we report both BLEU-4 and COMET~\cite{rei2020comet} to measure lexical translation accuracy and semantic translation quality, respectively.
BLEU-4 is computed at the corpus level using SacreBLEU~\cite{post2018call} with \texttt{13a} tokenization, while COMET is reported using the system-level score.
We use the \texttt{wmt22-comet-da}\footnote{https://huggingface.co/Unbabel/wmt22-comet-da} model for COMET.
Across both metrics, +T2L-ICAG consistently achieves the best results among the four systems. 
The gains are especially clear on BLEU-4, where T2L-ICAG improves the average score from 23.9 with +LoRA and 23.5 with +T2L-emb to 30.3. 
For COMET, T2L-ICAG also improves the average score to 0.775, compared with 0.721 for Kimi-Audio, 0.718 for +LoRA, and 0.732 for +T2L-emb. 
These results suggest that T2L-ICAG improves not only compositional task generalization in our main benchmark, but also multilingual translation transfer to unseen speech translation directions.

\begin{figure*}[t]
    \centering
    \includegraphics[width=\linewidth]{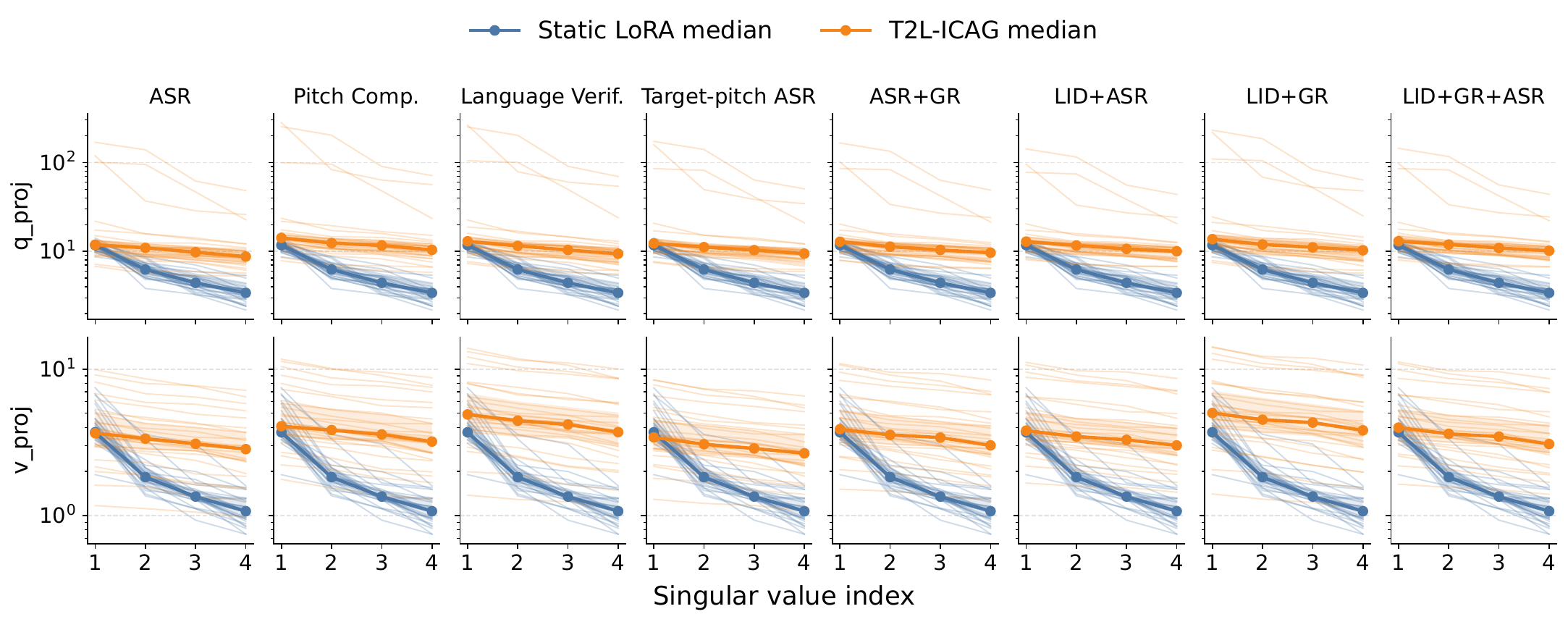}
    \caption{Singular-value spectra of static LoRA adapters and T2L-ICAG-generated adapters on Kimi-Audio. We only show the rank 4 results on 8 example tasks here, while the same pattern can be found in all tasks across all ranks. Light lines denote individual decoder layers, while dark lines denote median spectra over layers.}
    \label{fig:svd-spectrum}
\end{figure*}

\begin{table}[t]
\centering
\footnotesize
\renewcommand{\arraystretch}{1.06}
\setlength{\tabcolsep}{4pt}

\begin{tabularx}{\columnwidth}{>{\raggedright\arraybackslash}X*{2}{>{\centering\arraybackslash}X}}
\toprule
\textbf{Atomic} & \textbf{Acc. / Score\,$\uparrow$} & \textbf{WER\,$\downarrow$} \\
\midrule
\rowcolor{tablegray}ASR & -- & 1.21 \\
GR & 97.60 & -- \\
\rowcolor{tablegray}SER & 59.16 & -- \\
LID & 100.00 & -- \\
\rowcolor{tablegray}SQA & 89.66 & -- \\
\bottomrule
\end{tabularx}

\begin{tabularx}{\columnwidth}{l>{\centering\arraybackslash}X}
\toprule
\textbf{Comparative} & \textbf{Acc.\,$\uparrow$} \\
\midrule
\rowcolor{tablegray}Gender Verification & {57.73} \\
Duration Comparison & 96.75 \\
\rowcolor{tablegray}Pitch Comparison & 92.83 \\
Emotion Verification & 50.50 \\
\rowcolor{tablegray}Language Verification & 92.51 \\
\bottomrule
\end{tabularx}

\begin{tabularx}{\columnwidth}{l*{3}{>{\centering\arraybackslash}X}}
\toprule
\textbf{Selective execution} & \textbf{Acc.\,$\uparrow$} & \textbf{SAcc.\,$\uparrow$} & \textbf{mWER\,$\downarrow$} \\
\midrule
\rowcolor{tablegray}Target Gender LID & 95.65 & -- & -- \\
Target Language GR & 96.92 & -- & -- \\
\rowcolor{tablegray}Target Gender ASR & -- & 99.39 & 1.31 \\
Target Duration GR & 94.50 & -- & -- \\
\rowcolor{tablegray}Target Duration ASR & -- & 97.71 & 2.09 \\
Target Duration SER & 53.79 & -- & -- \\
\rowcolor{tablegray}Target Duration LID & 94.55 & -- & -- \\
Target Pitch ASR & -- & 95.66 & 1.57 \\
\rowcolor{tablegray}Target Gender SQA & 98.06 & -- & -- \\
\bottomrule
\end{tabularx}

\begin{tabularx}{\columnwidth}{l*{3}{>{\centering\arraybackslash}X}}
\toprule
\textbf{Sequential execution} & \textbf{Acc.\,$\uparrow$} & \textbf{IFR\,$\uparrow$} & \textbf{WER\,$\downarrow$} \\
\midrule
\rowcolor{tablegray}2-ASR & -- & 99.27 & 1.26 \\
3-ASR & -- & 98.97 & 1.29 \\
\rowcolor{tablegray}2-GR & 75.80 & -- & -- \\
3-GR & 77.40 & -- & -- \\
\rowcolor{tablegray}2-LID & 98.62 & -- & -- \\
3-LID & 99.52 & -- & -- \\
\bottomrule
\end{tabularx}

{\renewcommand{\tabularxcolumn}[1]{m{#1}}%
\setlength{\tabcolsep}{3pt}%
\begin{tabularx}{\columnwidth}{>{\raggedright\arraybackslash}m{0.28\columnwidth}*{4}{>{\centering\arraybackslash}X}}
\toprule
\textbf{Multi-function} &
\shortstack[c]{\textbf{LID}\\\textbf{Acc.\,$\uparrow$}} &
\shortstack[c]{\textbf{GR}\\\textbf{Acc.\,$\uparrow$}} &
\textbf{IFR\,$\uparrow$} &
\textbf{WER\,$\downarrow$} \\
\midrule
\rowcolor{tablegray}ASR + GR & -- & 98.56 & 98.21 & 1.29 \\
LID + ASR & 100.00 & -- & 99.95 & 2.74 \\
\rowcolor{tablegray}LID + GR & 100.00 & 96.95 & -- & -- \\
LID + GR + ASR & 100.00 & 97.85 & 99.95 & 2.68 \\
\bottomrule
\end{tabularx}}

\caption{Results of Qwen3.5-Omni-Plus on the 29-task compositional evaluation suite.
All values are percentages. IFR denotes the strict JSON format instruction-following rate.
For selective ASR tasks, SAcc. and mWER columns report selection accuracy and minimum WER, respectively; for other selective tasks, Acc.\ reports the final task performance.}
\label{tab:qwen35-full-results}

\end{table}

\begin{table}[t]
\centering
\renewcommand{\arraystretch}{1.3}
\setlength{\tabcolsep}{4pt}
\begin{tabular}{L{0.72\columnwidth}c}
\toprule
\textbf{Hypernetwork Instruction} & \textbf{Accuracy} \\
\midrule
Please verify if the following two speakers have the same gender. & 87.96\% \\
\midrule
\graycell{There are two audio clips, one from a male speaker and one from a female speaker. Please identify the gender of the speaker who speaks for a longer duration (Male or Female).} & \graycell{82.61\%} \\
Please recognize the gender of the following two speech clips respectively. & 76.74\% \\
\graycell{Identify the gender of the speaker.} & \graycell{75.19\%} \\
There are two audio clips in the following. Please transcribe only what the speaker with a higher average pitch said. Only output the transcript, without any other text. & 71.77\% \\
\graycell{Please transcribe the following audio:} & \graycell{65.73\%} \\
\midrule
The purple telescope quietly folds seventeen turquoise calendars under a whispering subway. & 63.21\% \\
\bottomrule
\end{tabular}
\caption{Accuracy of T2L-ICAG on the gender verification task across different hypernetwork instruction variants. 
Note that the LLM instruction is still the original task-inherent instruction.}
\label{tab:multi-instruction-accuracy}
\end{table}

\subsection{Case Study: Using Mismatched Instructions}
Usually, the instruction provided to the ICAG hypernetwork and the LLM backbone should be the same. 
In this section, we provide a case study on the gender verification task when such two instructions mismatch.
For the LLM instructions, we still use the task-inherent instruction, ``Please verify if the following two speakers have the same gender'', while for the ICAG hypernetwork, we experiment with instructions from other tasks and a random meaningless string.

The results in Table \ref{tab:multi-instruction-accuracy} show that all mismatched instructions lead to worse performance than the matched one. Specifically, performance decreases as the hypernetwork instruction deviates further from the original one.
This is a concrete example of T2L-ICAG learning to activate and strengthen corresponding capabilities of the LALM by generating LoRA adapters from instructions.

\subsection{Performance of State-of-the-Art LALMs}

\begin{figure}[t]
    \centering
    \includegraphics[width=\linewidth]
    {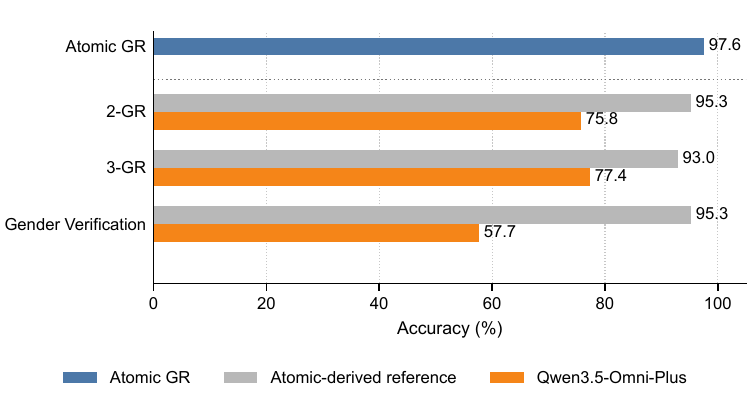}
    \caption{
    Gender-related compositional performance of Qwen3.5-Omni-Plus.
    Blue, gray, and orange denote atomic GR performance,
    atomic-derived expectations, and observed results, respectively.
    The expectations are diagnostic rather than strict upper bounds.
    }
    \label{fig:qwen35-gender-analysis}
\end{figure}

We additionally evaluate Qwen3.5-Omni-Plus~\cite{qwenteam2026qwen35omnitechnicalreport}, a strong proprietary
LALM, on our 29-task compositional benchmark using the same task
definitions and evaluation protocols. The model is tested through
direct inference without task-specific fine-tuning. We report its
task-native results in Table~\ref{tab:qwen35-full-results}.

Qwen3.5-Omni-Plus performs strongly on most tasks, achieving 1.21\% WER on atomic ASR, 100\% accuracy on language identification, and over 92\% accuracy on duration, pitch, and language comparison. It also performs well on most selective, sequential, and multi-function tasks. These results provide a useful sanity check that the constructed tasks are generally clear and solvable by a sufficiently capable LALM. Since its training data are not auditable, however, we do not interpret this performance as controlled evidence of compositional generalization. Instead, it serves as a reference point for attainable performance on our benchmark and highlights the remaining headroom for adapted open LALMs.

Despite its strong overall performance, Qwen3.5-Omni-Plus shows a
notable weakness on Gender Verification, achieving only 57.73\%
accuracy despite reaching 97.60\% on atomic gender recognition. Since the model also performs well on
Target Gender ASR, Target Gender LID, and other comparison tasks
involving duration, pitch, and language, the failure cannot
be attributed simply to weak gender perception or an inability to
compare two speech inputs. Rather, the relevant capabilities are
available individually but are not combined reliably under the Gender
Verification schema.

As shown in Figure~\ref{fig:qwen35-gender-analysis}, the model exhibits a
substantial gap between its atomic gender-recognition accuracy and its
gender-verification performance. A separate class-conditional breakdown on
the balanced verification set further reveals that the failure is highly
asymmetric: the model achieves 99.39\% accuracy on different-gender pairs but
only 16.07\% on same-gender pairs, indicating a strong bias toward the ``No''
prediction. These results show that even a highly capable LALM may fail to
compose familiar capabilities under a simple task schema. Compositional
generalization therefore remains imperfect and continues to be an important
direction for future work.

\subsection{Empirical Observations on Singular Values}
Figure \ref{fig:svd-spectrum} compares the singular-value spectra of static LoRA adapters and T2L-ICAG-generated adapters.
Specifically, for layer $i$ and target module $m$, we analyze the scaled LoRA update matrix $\Delta W_{i,m}$.
For each $\Delta W_{i,m}$, we compute
\begin{equation}
    \Delta W_{i,m}
    = U_{i,m}\operatorname{diag}(\sigma_{i,m,1},\ldots,\sigma_{i,m,r})V_{i,m}^{\top},
\end{equation}
and plot the singular values $\{\sigma_{i,m,k}\}_{k=1}^{r}$ in descending order.
For space reasons, we visualize Kimi-Audio rank-4 adapters on eight example tasks, while similar trends are observed in other rank settings.

Several empirical patterns can be observed.
First, the ICAG-generated adapters generally have larger singular values than static LoRA across the four singular values.
Second, their spectra are flatter: the later singular values remain relatively large, whereas static LoRA spectra decay more quickly.
Third, this pattern appears consistently across all visualized tasks and both target modules.
Finally, the generated adapters show visibly larger layer-wise dispersion, reflected by the broader spread of the light orange curves compared with the light blue curves.
These observations suggest that generated adapters have a different spectral profile from directly trained multi-task LoRA adapters.
We treat this as an empirical observation of instruction-conditioned adapter generation rather than a causal explanation of the performance gains.
Further investigation of the mechanism behind these spectral differences is left to future work.

\subsection{Consistency across LLM Judges}
To assess the reliability of our LLM-as-a-judge evaluation, we measure the inter-annotator agreement between the three LLM judges (Qwen3-32B, DeepSeek-v4-flash, GPT-4o-mini) on a random subset of the compositional benchmark.
We randomly choose 100 samples from each of 20 binary LLM-as-a-judge tasks from the three candidates: original Kimi-Audio, w/ LoRA ($r=4$) and w/ T2L-ICAG.
We report Krippendorff's $\alpha$~\cite{krippendorff2011computing} and the unanimous agreement rate for each system, as well as the overall agreement across all systems.
As shown in Table~\ref{tab:llm-judge-consistency}, three  LLM judge achieves strong agreement, with an overall Krippendorff's $\alpha$ of $0.935$ and a unanimous agreement rate of $95.8\%$, indicating that the LLM-as-a-judge protocol is highly reliable for our evaluation.

\begin{table}[t]
\centering
\renewcommand{\arraystretch}{1.12}
\setlength{\tabcolsep}{6pt}
\begin{tabular}{lccc}
\toprule
\textbf{System} & \textbf{N} & \textbf{Krippendorff's $\alpha$} & \textbf{Agree.} \\
\midrule
Kimi-Audio          & 2000 & 0.965 & 97.4\% \\
w/ LoRA       & 2000 & 0.894 & 94.8\% \\
w/ T2L-ICAG & 2000 & 0.888 & 95.3\% \\
\midrule
Overall           & 6000 & 0.935 & 95.8\% \\
\bottomrule
\end{tabular}
\caption{Inter-annotator agreement between the three LLM judges on a random subset of the compositional benchmark. ``N'' is the sample count, and ``Agree.'' denotes the fraction of samples on which all annotators agree.}
\label{tab:llm-judge-consistency}
\end{table}

\section{LLM Judge Prompts}

In this section, we provide the LLM-as-a-judge prompts used in our evaluations.
Placeholders such as \texttt{\{instruction\}}, \texttt{\{reference\}}, \texttt{\{response\}}, \texttt{\{prediction\}}, and \texttt{\{question\}} are filled in at evaluation time.
Unless otherwise noted, the judge returns a JSON object with a score and a short textual justification.

Figure~\ref{fig:evalsift-judge} shows the binary correctness prompt used for EvalSIFT, adapted from the official SIFT evaluation protocol~\cite{pandey2025sift}.

For our compositional tasks, we use a shared judge template and only change the task description.
Figure~\ref{fig:compositional-judge} shows two representative prompts:
(1) pitch comparison, which follows the shared template, and
(2) language identification, which adds explicit judging rules so that verbose paraphrases and unambiguous language synonyms are not penalized.
Other compositional tasks follow the same format.

\begin{figure*}[t]
\begin{llmprompt}
%%%%%%%%%%%%%%%%%%%%%%%%%%%%%%%%%%%%%%%%%%%%%%%%%%%%%%%%%%%%%%%%
%%% Example 1: Pitch Comparison (shared compositional template)
%%%%%%%%%%%%%%%%%%%%%%%%%%%%%%%%%%%%%%%%%%%%%%%%%%%%%%%%%%%%%%%%
You are asked to judge whether the output of a large audio language model is correct based on a ground truth reference. The large audio language model is supposed to complete a task described by a provided instruction.
Below is a question and the correct/reference answer about a pitch comparison task which is to judge if the first or the second speech input has a higher average pitch:
{ "question": "{instruction}", "answer": "{reference}" }
Rate the following answer as correct or not. If the above answer is correct, "score" should be 1, else 0. Note that if the model refuses to perform the task, the answer should also be regarded as incorrect.
Answer: {response}
Your output should be a JSON dictionary with keys as "reason" and "score". Do not include preambles, or any additional content.


%%%%%%%%%%%%%%%%%%%%%%%%%%%%%%%%%%%%%%%%%%%%%%%%%%%%%%%%%%%%%%%%
%%% Example 2: Language Identification (with synonym / paraphrase rules)
%%%%%%%%%%%%%%%%%%%%%%%%%%%%%%%%%%%%%%%%%%%%%%%%%%%%%%%%%%%%%%%%
You are asked to judge whether the output of a large audio language model is correct based on a ground truth reference. The large audio language model is supposed to complete a task described by a provided instruction.
Below is a question and the correct/reference answer about a single-clip spoken-language identification task:
{ "question": "{instruction}", "answer": "{reference}" }
Rate the following model answer as correct or not.
Judging rules:
- The reference answer is a single language label (for example, "English", "French", "German", "Italian", or "Spanish").
- Treat the answer as correct (score=1) if the model clearly identifies the same spoken language as the reference, even when the answer is verbose or paraphrased (for example, "The speaker is speaking English", "This audio is in French", or "The language spoken is German").
- Accept common equivalent names for the same language when unambiguous (for example, "Mandarin" for "Chinese", or "Castilian" for "Spanish"), but do not accept a different language.
- If the model mentions locale, country, or region instead of a language name, you may still score it as correct when that locale/country/region unambiguously implies the reference language (for example, answering "from France" for reference "French").
- Ignore extra polite text or brief explanation if the final language judgment is still clear and matches the reference.
- If the model refuses to answer, gives no language judgment, names multiple conflicting languages, or identifies a different language, score=0.
Model answer: {response}
Your output must be a JSON dictionary with keys "reason" and "score". "score" must be either 0 or 1. Do not include preambles or any additional content.
\end{llmprompt}
\caption{LLM judge prompts for our compositional tasks. Example~1 uses the shared template (pitch comparison); Example~2 adds synonym-aware judging rules (language identification).}
\label{fig:compositional-judge}
\end{figure*}

For spoken question answering, we adopt two AudioBench-style~\cite{wang2025audiobench} judge prompts.
Figure~\ref{fig:slue-sqa-judge} shows the 0--5 rating prompt used for SLUE SQA, and
Figure~\ref{fig:audiobench-mc-judge} shows the binary 0--1 rating prompt used for AudioBench multiple-choice questions (i.e., target-gender SQA on CN-College-Listen).

\begin{figure*}[t]
\begin{llmprompt}
Below is a question and the correct/reference answer about an audio:
{ "question": "{instruction}", "answer": "{reference}" }
Rate the following answer as correct or not. If the above answer is correct, "score" should be 1, else 0.
Answer: {response}
Your output should be a JSON dictionary with keys as "reason" and "score". Do not include preambles, or any additional content.
\end{llmprompt}
\caption{LLM judge prompt for EvalSIFT binary correctness evaluation.}
\label{fig:evalsift-judge}
\end{figure*}

\begin{figure*}[t]
\begin{llmprompt}
[Reference Answer]
{reference}

[Model Answer]
{prediction}

[Question]
{question}

[Task]
Rate the model's answer based on its alignment with the reference answer, focusing on accuracy and relevance to the reference provided. Please be critical on the details. If the model response is something like 'cannot decide', please rate as 0.
Criteria: Assess if the model's response mirrors the reference in terms of content, accuracy, and relevance.
Score0: The answer is refusing to give concrete results, providing something like 'cannot decide'.
Score0: The answer is completely misaligned, providing incorrect or irrelevant information compared to the reference.
Score1: The answer shows minimal alignment, often misunderstanding or providing irrelevant details unrelated to the reference.
Score2: The answer recognizes the topic but diverges significantly from the reference in accuracy or relevance.
Score3: The answer aligns with the reference generally but lacks detail or precise accuracy in some aspects.
Score4: The answer is mostly accurate and relevant, closely following the reference but could be clearer or more detailed.
Score5: The answer is highly accurate, detailed, and matches the reference answer perfectly, capturing its essence and detail.

Output requirements:
1) Return a valid JSON object only (no markdown, no code fences, no extra text).
2) The JSON must contain exactly these keys:
   - "explanation": string
   - "rating": integer in [0, 1, 2, 3, 4, 5]
3) Keep "explanation" concise and explicitly compare reference vs model answer.
\end{llmprompt}
\caption{LLM judge prompt for SLUE SQA (0--5 rating).}
\label{fig:slue-sqa-judge}
\end{figure*}

\begin{figure*}[t]
\begin{llmprompt}
[Reference Answer]
{reference}

[Model Answer]
{prediction}

[Question]
{question}

[Task]
Rate the model's answer based on its alignment with the reference answer, focusing on accuracy and relevance to the reference provided. Please be critical on the details.
Criteria: Assess if the model's response mirrors the reference in terms of content, accuracy, and relevance.
Score0: The answer is refusing to give concrete results, providing something like 'cannot decide'.
Score0: The answer is wrong, providing incorrect or irrelevant information compared to the reference.
Score1: The answer is correct, capturing or covering the meaning from the reference.

Output requirements:
1) Return a valid JSON object only (no markdown, no code fences, no extra text).
2) The JSON must contain exactly these keys:
   - "explanation": string
   - "rating": integer in [0, 1]
3) Keep "explanation" concise and explicitly compare reference vs model answer.
\end{llmprompt}
\caption{LLM judge prompt for AudioBench multiple-choice evaluation (binary 0--1 rating).}
\label{fig:audiobench-mc-judge}
\end{figure*}

\renewcommand{\refname}{Reference}
\bibliography{aaai2027}
\end{document}